\documentclass[ reprint,superscriptaddress, amsmath,amssymb, aps, prl]{revtex4-1}

\usepackage{graphicx}
\usepackage{dcolumn}
\usepackage{bm}
\usepackage[alsoload=synchem,alsoload=abbr]{siunitx}

\begin{document}

\title{Resonant X-Ray Holographic Imaging of the\\Insulator-Metal Phase Transformation in VO$_2$}

\author{L. Vidas}  
\email{luciana.vidas@icfo.eu}
\affiliation{ICFO - Institut de Ci\`encies Fot\`oniques, The Barcelona Institute of Science and Technology, 08860 Castelldefels (Barcelona), Spain}
\author{C. M. G\"unther}
\affiliation{Institut f\"ur Optik und Atomare Physik, Technische Universit\"at Berlin, 10623 Berlin, Germany}
\author{T. A. Miller}
\affiliation{ICFO - Institut de Ci\`encies Fot\`oniques, The Barcelona Institute of Science and Technology, 08860 Castelldefels (Barcelona), Spain}
\author{B. Pfau}
\affiliation{Max-Born-Institut, 12489 Berlin, Germany}
\author{D. Perez-Salinas}
\affiliation{ICFO - Institut de Ci\`encies Fot\`oniques, The Barcelona Institute of Science and Technology, 08860 Castelldefels (Barcelona), Spain}
\author{E. Mart\'inez}
\affiliation{ICFO - Institut de Ci\`encies Fot\`oniques, The Barcelona Institute of Science and Technology, 08860 Castelldefels (Barcelona), Spain}
\author{M. Schneider}
\affiliation{Max-Born-Institut, 12489 Berlin, Germany}
\author{E. Guhers}
\affiliation{Institut f\"ur Optik und Atomare Physik, Technische Universit\"at Berlin, 10623 Berlin, Germany}
\author{P. Gargiani}
\affiliation{ALBA Synchrotron Light Source, E-08290 Cerdanyola del Vall\`es, Barcelona, Spain}
\author{M. Valvidares}
\affiliation{ALBA Synchrotron Light Source, E-08290 Cerdanyola del Vall\`es, Barcelona, Spain}
\author{R. E. Marvel}
\affiliation{Department of Physics and Astronomy, Vanderbilt University, Nashville, Tennessee 37235-1807, USA}
\author{K. A. Hallman}
\affiliation{Department of Physics and Astronomy, Vanderbilt University, Nashville, Tennessee 37235-1807, USA}
\author{R. F. Haglund Jr}
\affiliation{Department of Physics and Astronomy, Vanderbilt University, Nashville, Tennessee 37235-1807, USA}
\author{S. Eisebitt}
\affiliation{Institut f\"ur Optik und Atomare Physik, Technische Universit\"at Berlin, 10623 Berlin, Germany}
\affiliation{Max-Born-Institut, 12489 Berlin, Germany}
\author{S. Wall}
\email{simon.wall@icfo.eu}
\affiliation{ICFO - Institut de Ci\`encies Fot\`oniques, The Barcelona Institute of Science and Technology, 08860 Castelldefels (Barcelona), Spain}

\date{\today}

\begin{abstract}
We use resonant soft X-ray holography to image the insulator-metal phase transition in vanadium dioxide with element and polarization specificity and nanometer spatial resolution. In contrast to recent measurements which integrated over a large area, our results show that electronic correlations do not change during the insulator-metal transition. Instead, we observe that nanoscale defects modify the local strain environment triggering the nucleation and growth of additional insulating phases on the nanoscale. Our technique shows that resonant imaging of phase transition in correlated materials is vital for understanding the origins of their macroscopic properties. 
\end{abstract}

\maketitle

VO$_2$ is a prototypical correlated material in which electronic correlations and structural distortions play a strong role in the physics of the system. At approximately \SI{60}{\celsius}, VO$_2$ undergoes a first order phase transformation from monoclinic (M$_1$) insulating phase to rutile (R) metallic phase. An outstanding issue has been whether the structural distortion alone can account for the opening of the band gap, or if electronic correlations are additionally required. Theoretical evidence for strong electronic correlations comes from the fact that density functional theory (DFT) often fails to produce an insulating state for the M$_1$ structure and instead predicts a metallic state for the low temperature structure. Although calculations with more sophisticated functionals~\cite{Eyert2011} have shown that a band gap can be obtained from DFT without correlations, other issues occur~\cite{Grau-Crespo2012} and the debate is still open. 

Experimentally it is hard to separate the contributions of electronic correlations from the structural change. While high quality single crystals show that structural and electronic changes are concomitant and occur within \SI{0.01}{\kelvin}~\cite{Budai2013}, in some thin films the metallicity can precede the structural change~\cite{Kim2015,Tao2012,Laverock2014}. Observation of a metallic monoclinic state would point towards the role of correlations as the primary driving mechanism for the phase transition and, recently, Grey {\em et al.}~\cite{Gray2016} used soft X-ray spectroscopy of thin films to suggest that electronic correlations weaken before the structure changes. 

However, unlike in single crystals, phase transitions in thin films can be several degrees wide, indicating the possibility of a heterogeneous transformation and phase coexistence. This is a particular concern in VO$_2$, as it is known that, in ambient conditions, it is close to a solid-state triple point with a second monoclinic insulating state (M$_2$)~\cite{Park2013}. Due to the first order nature of the phase transition, these phases may also coexist~\cite{Wu2006,Liu2015,Tselev2010}. To date, there has been no direct imaging of the monoclinic metallic phase, which makes it difficult to understand how this state fits in the phase diagram with the known monoclinic insulating and rutile metallic phases. 

To address this point, we use polarization and wavelength resolved X-ray holography to image the insulator-to-metal phase transition in thin films of VO$_2$. In contrast to the work by Grey {\em et al.}~\cite{Gray2016}, we do not find any evidence for a weakening in electronic correlations before the structural transition. Instead, we find that two different phase transitions occur within the sample, M$_1\rightarrow$R and M$_1\rightarrow$M$_2\rightarrow$R in spatially different locations. These distinct pathways result from nanoscale defects that modulate the strain environment locally within the sample. Our results highlight the key role played by defects in the phase transition in VO$_2$ and the power of resonant holography for studying phase separation in correlated material on the nanoscale. 

\begin{figure}[t!]
\includegraphics[width=0.4\textwidth]{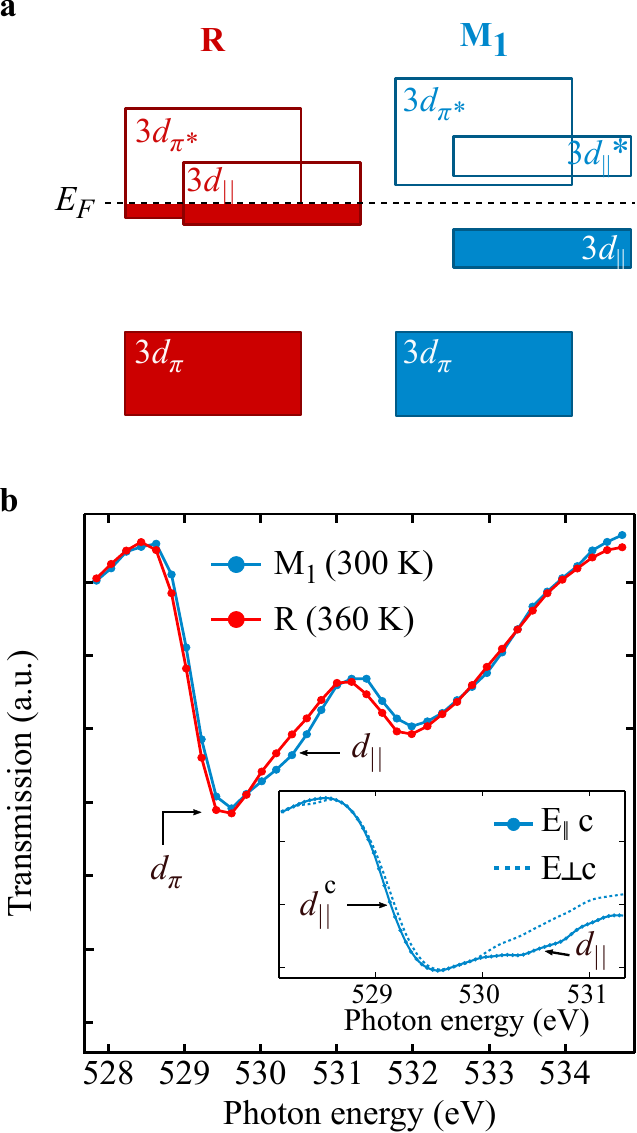}
\caption{\label{fig:fig1} Electronic structure and X-ray transmission of VO$_2$. (a) Schematic of the band structure proposed by Goodenough~\cite{Goodenough1971}. The $\pi$ states are hybridized vanadium $3d$ and oxygen $2p$ orbitals. The $d_{\parallel}$ state forms from overlapping vanadium $d$ orbitals along the rutile c-axis. (b) X-ray absorption spectra of the oxygen edge showing the transition to the $\pi^*$ and $d_{\parallel}$ states measured in transmission. The insert shows normalized spectra at room temperature, in the M$_1$ phase, for X-rays parallel and perpendicular to the rutile $c$-axis. The polarization-dependent edge shift, labelled $d_{\parallel}^c$, corresponds to the contribution from the $d_{\parallel}$ state due to the formation of correlated dimers\cite{Koethe2006,Biermann2005}.}
\end{figure}

Resonant soft X-ray holography is a lens-less imaging technique that can exploit the full power of element and polarization specificity of X-ray absorption spectroscopy (XAS) for contrast in imaging~\cite{Eisebitt2004}. A schematic band diagram for the states probed by XAS on VO$_2$ is shown in Fig.~\ref{fig:fig1}a, based on the Goodenough model~\cite{Goodenough1971}, for the M$_1$ and R phases. In this simplified picture, which does not include correlations, the $\pi$ levels are bonding/anti-bonding states that result from a hybridization of the vanadium $3d$ and oxygen $2p$ orbitals whereas the $d_{\parallel}$ states result from the overlap between vanadium ions along the rutile c-axis. The monoclinic distortion splits the $d_{\parallel}$ band and moves the $\pi^{*}$ orbital above the Fermi level. 

This simple model can qualitatively explain most of the XAS features. Spatially integrated XAS measurements on our VO$_2$ thin films on free-standing Si$_3$N$_4$ membranes are shown in Fig.~\ref{fig:fig1}b (see supplementary information for sample details). Due to the strong hybridization between the oxygen $2p$ and vanadium $3d$ levels, XAS spectra at the oxygen K-edge also probes the $3d$ orbitals of the vanadium ions. In the Goodenough picture, the bottom of the conduction band is the $\pi^{*}$ band, which is derived from V $d_{xy}$ and $d_{yz}$ orbitals. In the R state, the $\pi^{*}$ band moves below the Fermi level, resulting in an increase in absorption at lower energies, as observed. The $d_{\parallel}$ bands are derived from the V $d_{x^2-y^2}$ orbitals. Due to the strong anisotropy of these orbitals, transitions into these states are only observed when the electric field vector is parallel to the rutile c-axis. This is in contrast to the transitions into $\pi^{*}$ states, which are independent of the polarization. When the $d_{\parallel}$ band is split due to the structural transition in the low temperature phase, an additional higher energy absorption process is observed. This polarization-sensitive peak allows us to track both the structural transition and the orientation of the $c$-axis sample. 

Despite the fact that the $\pi^{*}$ states should be isotropic, the XAS shows a small, polarization-dependent, shift~\cite{Koethe2006} indicated as $d_{\parallel}^c$ in the insert of Fig.~\ref{fig:fig1}b. This cannot be explained by the Goodenough picture, but is reproduced by cluster dynamical mean field theory calculations~\cite{Biermann2005}, where an additional contribution to the bottom of the conduction band from $d_{\parallel}$ states occurs because of the formation of a vanadium singlet state as a result of electronic correlations. In the metallic phase, the dichroism in the $d_{\parallel}$ and $d_{\parallel}^c$ spectral features disappears due to the presence of transitions into all unoccupied V $d$ derived orbitals close to the Fermi energy in the absence of a bandgap.

Based on the temperature dependence of the XAS dichroism of the $d_{\parallel}$ and $d_{\parallel}^c$ states, Gray \textit{et al.}~\cite{Gray2016} argued that the singlet state is lost \SI{7}{\kelvin} below the conventional structural phase transition occurs and interpreted this as a sign of weakening electronic correlations prior to the structural transition. This may explain the origins for monoclinic-like metallic phases that have been observed~\cite{Kim2015,Tao2012,Laverock2014}. However, monoclinic metallic states are generally observed at temperatures in which both the metallic and insulating phases coexist and, from area-integrated XAS measurements, it is not clear if the loss of correlations is homogeneous or heterogeneous across the sample, or if nanoscale phase coexistence may affect the interpretation of the spectra. With resonant soft X-ray holographic imaging we can directly address this issue. 

\begin{figure}[t]
\includegraphics[keepaspectratio,width=0.45\textwidth]{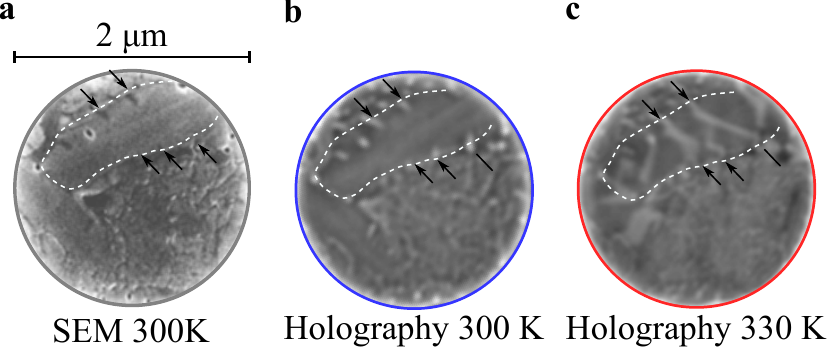}
\caption{\label{fig:fig2} (a) SEM images of the VO$_2$ sample show that large single crystals grow together with nanoscale crystallites. The dotted white line indicates the perimeter of a single crystal. On the edges of the crystal, several defects are observed (some indicated with arrows). (b) Holographic image of the same sample imaged at 530.5\,eV at room temperature, showing good agreement with the SEM image. (c) Same sample heated to 330\,K, i.e., close to the transition temperature (T$_c \approx$ 340\,K), thin stripes appear spanning the single crystal corresponding to the metallic phase.}
\end{figure}

Fig.~\ref{fig:fig2}a shows a scanning electron microscopy (SEM) image of a VO$_2$ sample. Although our substrate is amorphous, we occasionally find regions of the sample with micron-sized single crystals amongst smaller crystallites, and we focus on those in our study as indicated in Fig.~\ref{fig:fig2}. On the perimeter of these single crystals, several defects can be observed, with features as small as \SI{50}{\nano\metre}. In Fig.~\ref{fig:fig2}b, we image the same sample via X-ray holography with X-rays tuned to the $d_{\parallel}$ peak, which provides maximum contrast for the insulator-metal transition. At room temperature, we see excellent agreement between the SEM and holographic images given the different contrast mechanism of both techniques, demonstrating the adequate spatial resolution of our method. When we heat the sample to 330\,K, i.e., to the start of the phase transition in our films, we see bright stripes appearing across the crystallite (Fig.~\ref{fig:fig2}c). These stripes are metallic filaments that are observed because the sample has a higher transmissivity when metallic at the photon energy corresponding to the $d_{\parallel}$ XAS feature.  Interestingly, these domains, while long, can be as narrow as 50\,nm. It is clear from the evolution with temperature that the nanoscale defects nucleate the metallic phase and form metallic domains that span between defects, demonstrating the role of local strain in locally lowering the transition temperature.

We now turn to investigate the role of electronic correlations and the presence of other phases in the growth of the metallic phase at the insulator-metal transition. To do this, we perform spectrally resolved imaging at multiple X-ray wavelengths. The X-rays are linearly polarized along the rutile $c$-axis, as evidenced in the X-ray dichroic contrast shown in Supplementary Fig.~1. We collect X-ray holograms at 518\,eV incident photon energy on the vanadium L$_2$-edge, which is also sensitive to the $d_{\parallel}$ states~\cite{Haverkort2005}, at 529\,eV to be sensitive to the $\pi^*$ and $d_{\parallel}^c$ levels and at 530\,eV to be resonant to transitions into the $d_{\parallel}$ states. We then use the images obtained at these three photon energies to encode the red (518\,eV), blue (529\,eV) and green (530\,eV) channels of a false-color image. In this spectral fingerprinting approach, we can observe if changes occur in different regions of the sample and at different temperatures. For example, if electronic correlations precede the structural transition as observed by Gray \textit{et al.}~\cite{Gray2016}, then the $d_{\parallel}^c$ (blue channel) will change before the $d_{\parallel}$ (green channel), resulting in a change of color in the image. 

\begin{figure}[t]
\includegraphics[keepaspectratio,width=0.4\textwidth]{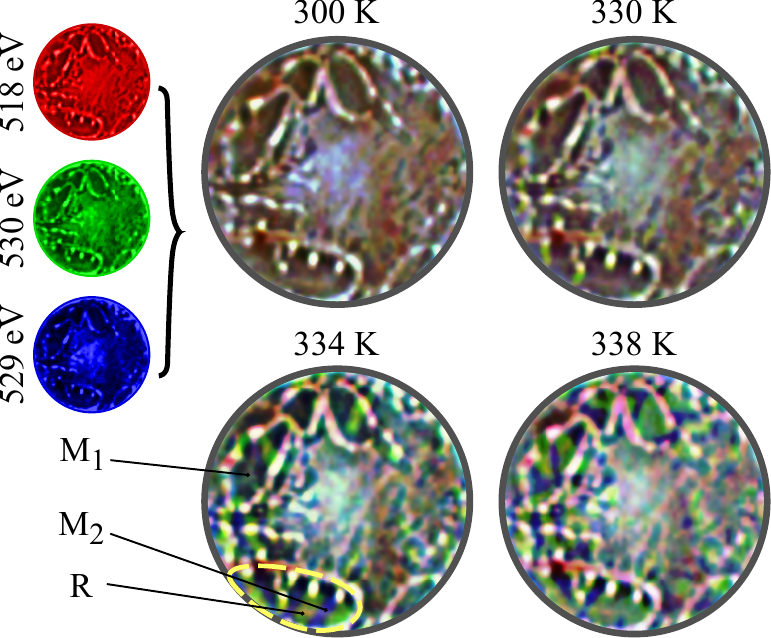}
\caption{\label{fig:fig3} Spectrally resolved images at the vanadium and oxygen edges. The images are recorded at 518, 529 and 530\,eV and are used to encode the intensities of the three color channels of an RGB image. At 330\,K, an increase in intensity of the green channel, which probes the rutile phase through the $d_{\parallel}$ state, is observed in small regions similar to that observed in Fig.~\ref{fig:fig2}. As the sample is heated further, it becomes increasingly clear that the blue channel, which probes the $d_{\parallel}^c/d_{\pi^*}$ state, also changes but in different regions. At 334\,K, three distinct regions can be observed corresponding to the monoclinic M$_1$, M$_2$ and R phases. As the temperature increases, the R phase dominates. Dashed line corresponds to the ROI shown in Fig.~\ref{fig:fig4}. The circular field of view is \SI{2}{\micro\metre} in diameter.}
\end{figure}

Fig.~\ref{fig:fig3} shows the spectrally resolved changes as a function of temperature in a different sample to that shown in Fig.~\ref{fig:fig2}, the SEM image of which is shown in Supplementary Fig~2. Again, we see the formation of nanoscale domains that span the small single crystals, starting at defects. However, now we can see that in different regions of the sample the X-ray transmission changes differently. At 334\,K, we can discern at least three different phases as indicated: the original M$_1$ phase and two new phases, one with a subtle change in the $d_{\parallel}^c$ (blue) channel and a second one with the greater change in the $d_{\parallel}$ (green) channel.

\begin{figure}[t!]
\includegraphics[width=0.35\textwidth]{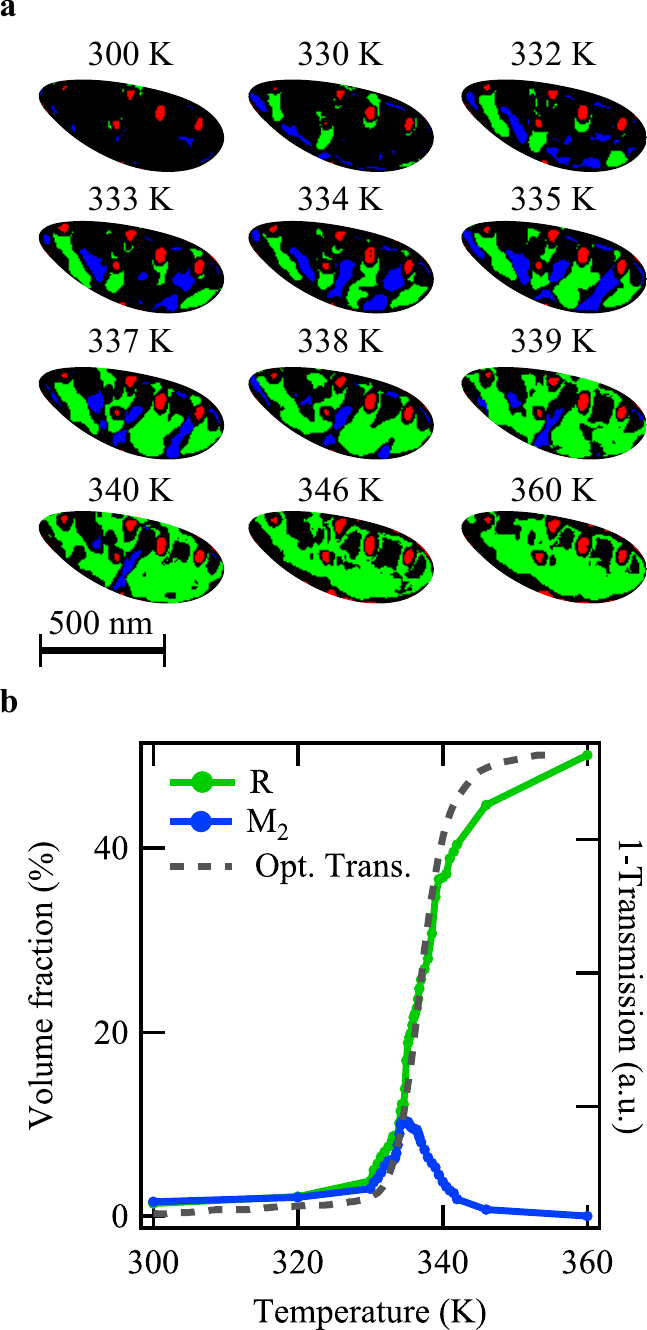}
\caption{\label{fig:fig4} (a) Threshold images of domain growth in VO$_2$ of the single crystal outlined in Fig.~\ref{fig:fig3} at selected temperatures. Growth starts from defects (red) and consists of two phases M$_2$ (blue) and R (green). (b) Temperature dependent growth of the two phases, the volume fraction of the M$_2$ phase peaks at $\sim$335\,K before the whole sample becomes metallic. The growth of the R phase is in good agreement with the change of the optical transmission measured on a witness sample.}
\end{figure}

To understand how these phases grow, we perform a finer temperature scan and focus on one of the small single crystals indicated in Fig.~\ref{fig:fig3}. In addition, to amplify the changes we threshold each channel (see supplementary information) and show images at key temperatures in Fig.~\ref{fig:fig4}a. At low temperature, the sample is predominantly black within this color-coding scheme, indicating the presence of the initial M$_1$ phase in the crystallites, together with some defects (red). As the sample is heated, the new phases (blue, green) nucleate at the boundaries and defects of the single crystals. At intermediate temperatures (334\,K), a striped state is present between regions with changes primarily in the $d_{\parallel}^c$ channel and those in the $d_{\parallel}$ channel, before the changes tracked by the $d_{\parallel}$ feature dominate the region. This phase growth is captured in Fig.~\ref{fig:fig4}b, which shows the volume fraction of the blue and green regions as a function of temperature within the region of interest (ROI). Both phases start to nucleate at similar temperatures and grow, before the green phase dominates. 

The green, $d_{\parallel}$, domains can be easily interpreted as the metallic R phase, that dominates at high temperatures. If changes in electronic correlations precede the structural transition, we would expect any green regions to first turn blue as changes at $d_{\parallel}^c$ should occur before those at $d_{\parallel}$~\cite{Gray2016}. However, we note that many regions show a direct transition from M$_1$ to R without weakening of correlations. Only in some spatially distinct regions we do observe the intermediate blue phase. This indicates that the loss of correlations is not the driving mechanism for the insulator-metal phase transition as previously reported. Instead, we suggest that the blue phase can be assigned to the insulating M$_2$ phase, which can appear in strained samples. This assignment is justified by recent calculations of the M$_2$ phase density of states, which show changes in the $d_{yz}/d_{xz}$ orbitals~\cite{Quackenbush2016}, which would also result in changes in the XAS in the vicinity of $d_{\parallel}^c$. 

These spatially resolved measurements allow us to draw a new interpretation of the phase transition in thin films. Nanoscale defects in the thin film modify the local strain environment, locally reducing the phase transition temperature for M$_1\rightarrow$ R. Due to the large volume difference of the R phase, a new strain field is generated which can also nucleate the M$_2$ phase. Both phases continue to grow and form a striped phase due to the interaction of the strain fields, as also observed in larger nanobeam single crystals\cite{Wu2006,Liu2015,Liu2014,Jones2010}. As the temperature is further raised, a complete transformation of the M$_2$ to R occurs.

This interpretation can explain the usual temperature dependence of the XAS signal in the vicinity of $d_{\parallel}^c$ during a thermal cycling previously reported~\cite{Ruzmetov2007}, as the M$_2$ phase can be stabilized back to room temperature. Furthermore, while the features at $d_{\parallel}^c$ may result from electronic correlations, changes in this region can also be due to structural transitions. As a result, we do not find any evidence for a new monoclinic metallic phase of VO$_2$ nor do we find evidence for a weakening of electronic correlations in the insulator-metal phase transition. Instead, we find that the phase transition can be explained in terms of nanoscale phase separation into the known phases of VO$_2$. In some respects, observation of the M$_2$ phase in our samples is surprising as there is no epitaxial strain provided by the substrate to move the sample away from the M1$\rightarrow$R pathway. However, this work demonstrates the role defects play in locally modifying the strain environment which can dramatically modify the phase transition in correlated materials. As a result, we have shown that resonant holography can play a vital role in understanding the XAS spectra and transition pathway in correlated materials with nanoscale defects.\\

\begin{acknowledgments}
T.A.M.\ and L.V.\ thankfully acknowledge financial support by the HZB. S.W.\ acknowledges financial support from Spanish MINECO (Severo Ochoa grant SEV-2015-0522), Ram\'on y Cajal programme RYC-2013-14838, Marie Curie Career Integration Grant PCIG12-GA-2013-618487, Fundaci\'o Privada Cellex, and CERCA Programme / Generalitat de Catalunya.  Research at Vanderbilt was supported by the United States National Science Foundation (EECS-1509740 for KAH, DMR-1207507 for REM). We thank Frank de Groot and Frederica Frati for insightful discussions.  
\end{acknowledgments}

\renewcommand\thefigure{S\thesection.\arabic{figure}}   
\section{Supplementary Information}
\setcounter{figure}{0}    
\subsection{\label{app:sample}Sample fabrication}

VO$_2$ samples, \SI{75}{\nano\metre} thick, were deposited, using pulsed laser deposition (PLD), onto Si$_3$N$_4$ membranes featuring a \SI{1.2}{\micro\metre} thick Cr(\SI{5}{\nano\metre})/Au(\SI{55}{\nano\metre}) multilayer layer on the reversed side. Prior to deposition, the chamber of an Epion PLD-3000 system was pumped down to \SI{90}{\mu Torr} and the vanadium metal target was ablated to remove surface contamination. The deposition was performed at pressure of \SI{11}{\torr} while maintaining a \SI{2}{sccm} flow of ultra-high purity oxygen gas. The sample was \SI{8}{\centi\metre} from the target during deposition. The VO$_2$ thin films were deposited by ablating the target with a KrF excimer laser (wavelength \SI{248}{\nano\metre}). The beam (\SI{4}{\joule\per\square\centi\metre} per pulse, \SI{25}{\Hz} repetition rate, and \SI{25}{\nano\second} pulse duration) was rastered across the rotating vanadium metal target during deposition. After deposition, the samples were annealed in a tube furnace at \SI{450}{\celsius}, while the O$_2$ flow rate (19–22 sccm) was actively adjusted to maintain \SI{250}{\milli\torr}.
After deposition, the gold multilayer was used to define the holography mask. A focused ion beam was used to mill an aperture of \SI{2}{\micro\metre} in diameter through the multilayer to enable transmission of the X-rays. Additionally, three reference holes with exit diameters of 50–90 nm, were drilled through the gold, Si$_3$N$_4$ and VO$_2$ layers. 

\subsection{\label{app:x-rays}X-ray measurements}

Measurements were carried out at the UE52-SGM beamline of the BESSY II synchrotron radiation source at Helmholtz Zentrum Berlin using the ALICE X-ray scattering instrument\cite{Abrudan2015}. The instrument is equipped with a cryostat and resistive heaters which were used to control the sample temperature with a stability of \SI{0.1}{\kelvin}. The charge-coupled device (CCD) camera was placed approximately \SI{40}{\centi\metre} away from the sample. This geometry corresponds to a maximum detectable in-plane momentum transfer of about \SI{90}{\per\micro\metre} and a maximum spatial resolution of \SI{41}{\nano\metre}. XAS was measured via transmission of the thin film in the areas not closed by the gold mask with the incident X-ray beam parallel to the sample normal. Fourier transform holograms were recorded in the same transmission geometry.  All images shown are taken from the real part of the sample’s exit wave as reconstructed by taking the Fourier transformation of the hologram. A beam block was used to block the highly intense central beam in order to adapt the dynamic range of the scattering pattern to the capabilities of the detector.

The beam block acts as a low pass filter and essentially subtracts the average transmission of the entire sample. As a result, the absolute transmitted intensity for each phase depends on the average phase of the whole sample, which is changing as the phase fraction grows. However, although the absolute values change, the contrast between phases remains. As a result, the threshold used in Fig.~4 is manually adjusted to ensure the contrast is not lost for each temperature. In addition, an upper threshold is applied to the blue and green channel in order to move the white regions, corresponding to defects and boundaries of the crystals seen in Fig.~3, into the red channel in Fig.~4. Polarization-dependent images and spectra were measured in the MaReS end station at BOREAS beamline at the ALBA synchrotron\cite{Barla2016}.

\begin{figure}[t!]
\includegraphics[width=0.4\textwidth]{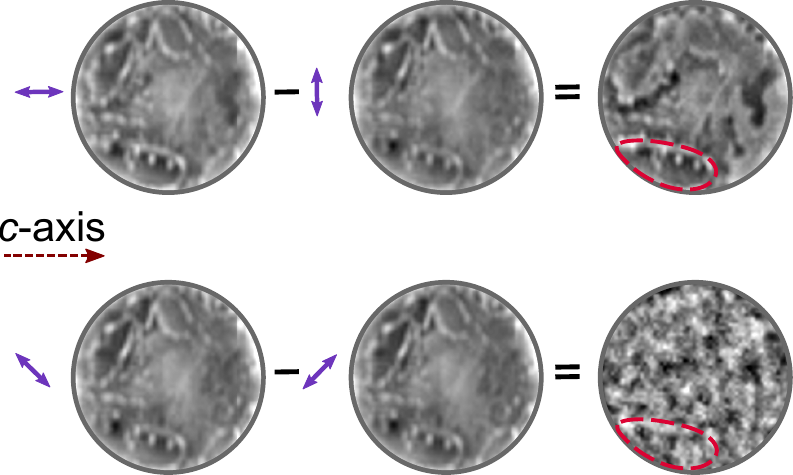}
\caption{\label{fig:Sfig1} Polarization resolved imaging on VO$_2$.  X-ray dichroic images of VO$_2$ films measured at the $d_{\parallel}$ peak (\SI{530}{\electronvolt}) at room temperature. The X-ray polarization was rotated by 180 degrees in 5 degree steps (not shown). When the X-rays are parallel/perpendicular to the rutile $c$-axis, the dichroic (difference) image is strongest due to the anisotropy of the $d_\parallel$ state (top row). This difference is lost when each polarization is rotated by 45 degrees with respect to the $c$-axis (bottom row). The dashed red line corresponds to the ROI used in Fig.~3 and 4. Field of view is 2\SI{2}{\micro\metre} in diameter.}
\end{figure}

\begin{figure}[t!]
\includegraphics[width=0.3\textwidth]{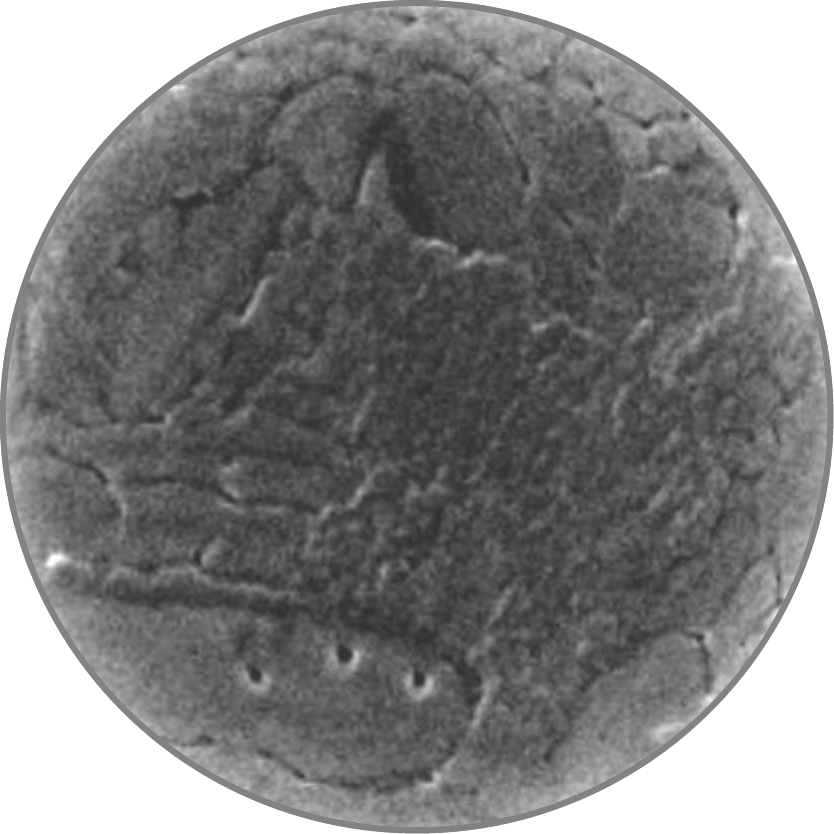}
\caption{\label{fig:Sfig2}SEM image of the second sample used in Fig.~3, 4 and Supplementary Figure~\ref{fig:Sfig2}. Field of view is \SI{2}{\micro\metre} in diameter.}
\end{figure}

\bibliography{HoloPRL}

\end{document}